\documentclass[psf,epsf,twocolumn,showpacs,preprintnumbers,floatfix,prb]{revtex4}
\usepackage{graphicx}
\usepackage{graphicx}
\usepackage{dcolumn} 
\usepackage{bm}
\usepackage{epsfig}
\begin{document} 

\title{Magnetism from $2p$ States in Alkaline Earth Monoxides:\\
        Trends with Varying N Impurity Concentration}

\author{V. Pardo$^{a,b}$ and W. E. Pickett$^{b}$}
\affiliation{$^{a}$Departamento de F\'{i}sica Aplicada, Facultad de F\'{i}sica,
  Universidad de Santiago de Compostela, E-15782 Campus Sur s/n,
  Santiago de Compostela, Spain}
\affiliation{$^b$Department of Physics,
  University of California, Davis, CA 95616}
\begin{abstract}
$2p$-based magnetic moments and magnetic coupling are studied with density
functional based methods for 
substitutional N in the alkaline earth monoxide series MgO, CaO, SrO,
BaO.  The hole is rather strongly localized near the N$^{2-}$ ion,
being somewhat more so when strong on-site interactions are included in
the calculations. Strong magnetic coupling is obtained in the itinerant electron limit but 
decreases strongly in the localized
limit in which the Coulomb repulsion within the N $2p$ shell (U) is much
greater than the N $2p$ impurity bandwidth (W).
We find that this limit is appropriate for realistic N concentrations. 
Ordering on a simple cubic sublattice may maximize the magnetic coupling due to its high directionality.
\end{abstract}
\maketitle

\section{Introduction}

After a strong research effort spent on magnetic-ion doped semiconductors
as spintronics materials,\cite{RMP} 
scientific effort has recently turned to the possibility of obtaining 
useful ferromagnetic (FM) order at room temperature in materials without conventional 
magnetic atoms with open $d$ or $f$ shells. 
Isolated defects in insulators often produce magnetic moments, which occur anytime
a state in the band gap is occupied by an odd number of electrons.  A common example
is the P impurity in Si, where the magnetism of the material evolves with increased
doping as the insulator-metal transition is approached. Such strictly localized 
states in insulators are magnetic regardless of their underlying atomic character 
or their degree of localization.

Elfimov et al. \cite{elf1} proposed to use the cation vacancy -- the charge conjugate of the
conventional F-center state -- in non-magnetic oxides such as 
CaO\cite{vacancy} as a source of magnetic moment in insulators. 
The resulting pair of holes, which are bound to 
the region for an isolated 
vacancy, may form a magnetic center even if it does not lie in the gap,
due to the large Hund's coupling on the 
oxygen ions neighboring the vacancy (which can no longer be simple non-magnetic O$^{2-}$).  These states
extend to a few atomic neighbors, with magnetic coupling resulting from direct
exchange.  Magnetic order becomes possible, and
at concentrations of a few percent they suggested that conduction, and half 
metallic ferromagnetism, might result.

The Osaka group\cite{osaka0} followed by suggesting that B, C, or N 
substituted for O in CaO
could result in local moments in the $2p$ states of the impurity ions.  The 
coupling between moments was found, in calculations on random alloys, to be FM and
half-metallic ferromagnetism was also predicted, possibly as high as
room temperature.  Elfimov et al. \cite{elf2} studied the specific case of N substituting for
O in SrO, applying not only the conventional local spin density approximation (LSDA)
but also the strongly-correlated form (LSDA+U) that should be appropriate
for localized magnetic states.  They also obtained the possibility  of half metallic
ferromagnetism at technologically relevant temperatures.
Experimentally, magnetism has been 
observed as a surface effect in thin films of these non-magnetic oxides.\cite{hong}

\begin{table}[h!]
\caption{Lattice parameters\cite{osaka} and band gap\cite{band_gaps} of the corresponding monoxides under study.}\label{data_tab}
\begin{center}
\begin{tabular}{|c|c|c|}
\hline
Compound & Lattice parameter (\AA) & Band gap (eV) \\
\hline
\hline
MgO & 4.12 & 7.2 \\
CaO & 4.81 & 6.2 \\
SrO & 5.16 & 5.3 \\
BaO & 5.52 & 4.0 \\
\hline
\end{tabular}
\end{center} 
\label{table1}
\end{table}

There are other examples of atomic $p$ character giving rise to magnetism in
solids.  One example is the class of alkali hyperoxides (for example, 
RbO$_2$\cite{lines,rbo2} and Rb$_4$O$_6$\cite{rb4o6,degroot}).  These materials
are ionic, containing the spin-half O$^{1-}_2$ ion, and display a variety of
unusual magnetism-related properties.  The SrN$_2$ compound, an ionic material
containing the N$^{2-}_2$ unit, is a related example,\cite{auffermann}
while the SrN compound with magnetic N is predicted to be a half metallic
ferromagnet.\cite{volnianska}  Between the bulk compounds and the (nearly)
isolated magnetic impurities, another promising example has been reported.  At
the $p$-type interface between LaAlO$_3$ and SrTiO$_3$, there is (from simple
electron counting) 0.5 too few electrons per interface cell to fill all of the
O $2p$ states that are filled in both bulk materials.  Holes mush exist in the
O $2p$ bands at the interface, and Pentcheva and Pickett predicted that magnetic
holes arise and should order at low temperature.\cite{pentcheva}  Subsequently,
magnetic hysteresis has been observed at this interface.\cite{magn_IF}

Here we will analyze the magnetism and magnetic coupling in alkaline-earth 
monoxides doped with N in more detail.  Substitution of O
by N in an insulating host presents a clear platform for local moments, and
the basic picture is simple.  Besides
changing the potential locally, the
substitution of O by N removes one electron giving a N$^{-2}$ $2p^5$ ion, 
and thereby introduces a hole that is necessarily
magnetic (being spin-uncompensated).
Ferromagnetism  may result if the
magnetic coupling has the right character and is sufficiently robust.
Within the series MgO, CaO, SrO and BaO the volume varies 
by a factor of 2.3 and the bandgap varies by over a factor of 1.8, see Table \ref{data_tab}. 
By introducing dopant nitrogen atoms
with concentrations of one N atom per 8, 16 and 32 O atoms, 
we are able to model impurity concentrations of 12.5\%, 6\% and 3\%, 
respectively. The goal here is to understand the electronic structure and
magnetic character of the 
substitutional N impurity, to study dopant interactions
for the various concentrations, and to analyze the magnetic properties by
calculating the condition for strong coupling that could lead to a 
FM T$_c$ on the order of room temperature. 

These semiconductors crystallize in a rocksalt structure.  Upon  
increasing the lattice parameter by introducing a bigger cation, there should
be a tendency for the N hole state to become more {\it localized}.  However, the 
band gap narrows in proceeding from the smallest to largest
(7.2 eV for MgO, 4.0 for BaO), a trend that should
cause a corresponding increase in the tendency 
towards {\it delocalization} of the impurity state.  How these effects 
compete is important to determine.  Another distinction is that the 
lowest conduction states in CaO, SrO, and BaO are $d$ states, while MgO
is very different in this regard (Mg $s$, O $3s, 3p$).  The hole state
will be formed from valence orbitals, however, so the impact both of the
magnitude of the band gap and the character of the conduction band must
be calculated.

Table \ref{dist_tab} summarizes the N-N nearest neighbor distances in the different compounds 
for the various concentrations considered. As the concentration changes, 
the coordination of the impurities varies, as is provided in the table.

\begin{table}[h!]
\caption{Nearest neighbors distances and number of nearest neighbors for
the impurity at various concentrations and monoxides under study.}\label{dist_tab}
\begin{center}
\begin{tabular}{|c|c|c|}
\hline
Compound & Concentration & N-N distance (\AA) \\
& & (no. neighbors) \\
\hline
\hline
MgO & 1/8 &  6.0 (12)\\
MgO & 1/16 & 6.0 (4) \\
MgO & 1/32 & 8.4 (6) \\
CaO & 1/16 & 6.8 (4) \\
SrO & 1/8 &  7.3 (12) \\
SrO & 1/16 & 7.3 (4) \\
SrO & 1/32 &  10.3 (6) \\
BaO & 1/8 &  7.8 (12)\\
BaO & 1/16 & 7.8 (4)\\
BaO & 1/32 & 11.0 (6) \\
\hline
\end{tabular}
\end{center} 
\end{table}

\section{Computational details}

Electronic structure calculations were  performed within density functional 
theory\cite{dft} using {\sc wien2k},\cite{wien} which utilizes an augmented 
plane wave plus local orbitals (APW+lo)\cite{sjo} method to solve the Kohn-Sham 
equations. This method uses an all-electron, full-potential scheme that makes no shape 
approximation to the potential or the electron density. The exchange-correlation 
potential utilized was the Wu-Cohen version of the generalized gradient 
approximation\cite{wu_cohen} and strong correlation effects were introduced by means 
of the LSDA+U scheme\cite{sic} including an on-site effective U for the O and N p 
states. We have used option nldau= 1 in {\sc wien2k}, i.e. the so-called ``fully-localized limit", using an effective U$_{eff}$= U-J, being J the on-site Hund's rule coupling constant, taken as J= 0. The justification of utilizing the LSDA+U scheme comes from spectroscopic 
measurements in oxides, which estimate large values of U (5-7 eV) for the p 
states.\cite{auger_Up1} This has been shown to be crucial in understanding the interfacial electronic structure of transtion metal oxides, where magnetic moments can localize in the O atoms.\cite{pentcheva}
All the calculations were converged with respect to all the parameters
used, to the precision necessary to support our calculations.

\begin{figure}[ht]
\begin{center}
\includegraphics[width=0.9\columnwidth,draft=false]{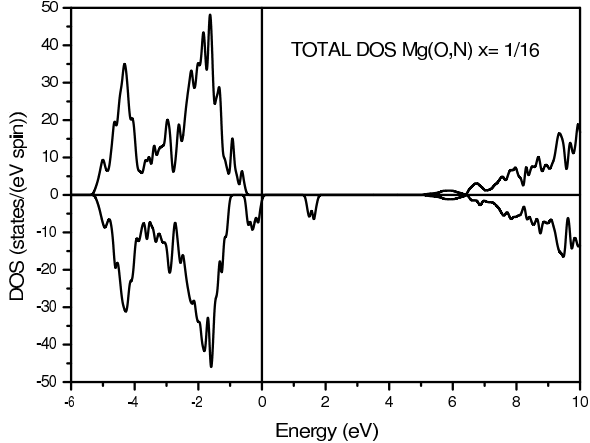}
\caption{Total density of states for ferromagnetically aligned
MgO$_{1-x}$N$_x$ with $x$= 1/16
$\approx$ 6\%, calculated with the LSDA+U method including spin-orbit
effects using U= 5.5 eV
for the $2p$-states of both N and O. The majority DOS is plotted upward,
the minority is plotted downward.}
\label{1DOS}
\end{center}
\end{figure}

\section{Results}

We begin by studying the local electronic structure of a N $2p^5$ impurity 
in MgO. Comparing a non-magnetic solution for the N atom 
and a solution with 1 $\mu_B$/N (a local moment formation in the N site), 
the total energy difference favors the magnetic solution by more than 
100 meV/N, a value that is nearly independent of the concentration of 
impurities.  
The introduction of a value of 
U$_{eff}$ = 5 eV produces a further stabilization of the magnetic solution 
with respect to the non-magnetic one (the energy difference rises above 
300 meV/N). 

\subsection{Effect of atomic relaxation}
Substitutional N
in these monoxides will break the symmetry and may
lead to some local distortion of the otherwise
perfectly octahedral environment of the cation. To determine the importance
of atomic relaxation we performed
structural minimization for MgO$_{1-x}$N$_x$, with $x$= 1/16.
If sufficiently localized, the hole might occupy 
a p$_z$ orbital (l$_z$= 0), leading to an elongation of the N-Mg distance
along the z axis,
or it could be p$_x$$\pm$$i$p$_y$ (l$_z$= $\pm$ 1) that is unoccupied,
leading to an elongation of the N-Mg distance in the plane perpendicular
to the z axis. The symmetric LSDA solution shares the hole among the p
orbitals equally.
We have performed a structural optimization within GGA
({\it i.e.} without including strong correlation effects) which leads to an electron
density describing the hole in the N atom being symmetric.
The result is that Mg-N distances get modified only slightly, 
becoming elongated by less than
0.5\%.  Also, using the LSDA+U scheme for doing the structure
optimization, the hole located in a p$_z$ state leads to
an elongation of about 3\% of the Mg-N distance along the z-axis.
We have found that these small differences have little effect on the trends we study, either the
localization of the magnetic moment or the magnetic coupling, hence
we have neglected relaxation in the results that follow.

\begin{figure}
\includegraphics[width=0.9\columnwidth,draft=false]{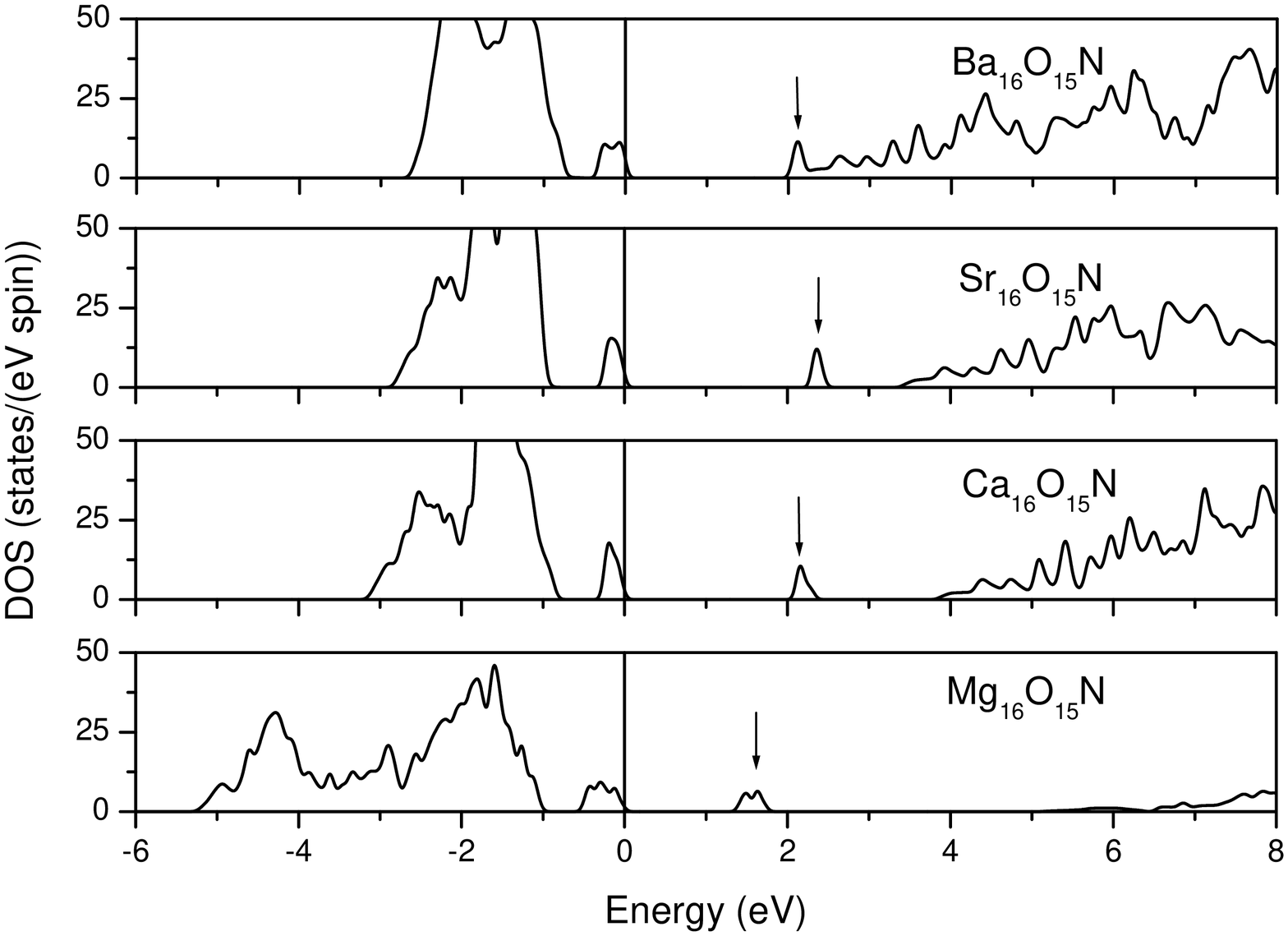}
\caption{Total density of states of the various monoxides under study for FM alignment,
calculated with the LSDA+U method for U= 5.5 eV. Only the minority spin
channel is shown, since it illustrates all the important features:
O $2p$ bandwidth, narrowing of the bandgap, and the position of the hole
state at 2-2.5 eV above the valence band maximum..}\label{all_dos}
\end{figure}

\subsection{Electronic and Magnetic Trends}
We have treated the N concentrations listed in Table \ref{dist_tab}. 
For each concentration, we have assumed that the N impurities are as regularly 
distributed as possible (their separation is maximized). As
mentioned above,
one can imagine how the physics can change from MgO 
to BaO as the size of the cation, and the volume, increases. 
On one hand, greater delocalization 
is permitted as the band gap is reduced 
(by a factor 2 experimentally, by a factor 3 from our calculations for the doped
compounds with U$_{eff}$= 5.5 eV). On the other hand, for a given concentration of
impurities, 
the N-N distance 
becomes larger for the larger volume compounds. The former could encourage an easier propagation 
of the magnetic 
coupling throughout the crystal, while the latter would 
imply a smaller magnetic 
interaction if the size of the defect state remains the same.

We first consider the U$_{eff}$= 5.5 eV electronic structure for a particular 
case, MgO with $x$= 1/16 $\sim$ 6\%, where the density of states (DOS) for separate spin
directions is shown in Fig. \ref{1DOS}.
For the minority spin, the narrow hole band lies in the gap and
has a full bandwidth of 
W $\approx$ 0.3 eV.  The other two N $2p$ minority states are occupied
and lie above the O $2p$ bands.  The separation of filled and unfilled
minority N states is 2 eV, being the Hund's exchange about 0.6 eV in the
occupied N p states. The conduction bands show no
appreciable spin splitting. When we include spin-orbit coupling, the hole occupies the
p$_x$ - $i$ p$_y$ state, l$_z$= -1 orbital, satisfying Hund's rule.

To display the trends across the series, it is sufficient to look
only at the minority DOS, which is plotted
for each compound in Fig. \ref{all_dos}. 
The band gap reduction through the series Mg$\rightarrow$Ba is evident.  
The O $2p$ bandwidth is much larger for MgO but this has little
effect on the defect state, except perhaps contributing a small additional broadening.
In spite of this large band gap variation, and the change in the volume
(hence the near neighbor distance) the position of the unoccupied hole state
changes rather little, staying 2$\pm$0.5 eV above the highest occupied state.
The variation that does occur is non-monotonic and may be due to the
conduction bands that fall in energy through the series.
For BaO, narrowest band gap member, the N hole band merges into the 
bottom of the conduction Ba d bands.

Because the intra-atomic Coulomb interaction strength U is at least a few eV,
these systems are in the regime U/W $>>$ 1 (being W the N 2p hole
bandwidth) so hopping of the holes 
will be inhibited, giving insulating behavior.
This is the limit in which the LSDA+U method works well.
For values of U bigger than $\sim$1 eV, the material is insulating whereas for 
smaller values (i.e. in LSDA or GGA), the N hole band intersects
the upper valence
bands and the band structure
becomes metallic (or half-metallic).

\subsection{Sign, strength, and character of magnetic coupling}

Since the N ions form a strong local magnetic moment, 
the next question is whether they couple ferromagnetically and what the 
strength of the magnetic interaction is.
We have performed calculations on both FM and antiferromagnetic (AF) 
alignments to obtain the coupling between nearest neighbor N moments, 
for the different 
concentrations and monoxides under study. We estimate the Curie temperature 
by using  the mean field expression from a Heisenberg model of the type H= - $\sum$ J$_{ij}$S$_i$S$_j$, with S=1/2 (the sum is over pairs),
\begin{eqnarray}
 T_c= \frac{2zS(S+1)J}{3k_B}, 
\end{eqnarray}
where $z$ is the number of nearest neighbors and $J$ is the exchange constant.

\begin{table}[ht]
\caption{Strength of the magnetic coupling converted into Curie
temperature of the corresponding monoxides
under study obtained from an LSDA calculation (including spin-orbit
effects) and a GGA (Wu-Cohen) calculation (U= 0).}\label{tc_tab}
\begin{center}
\begin{tabular}{|c|c|c|c|}
\hline
Compound & Concentration & T$_c$ (K)  & T$_c$ (K) \\
         &               &  [LSDA+SO]  &  [GGA]   \\
\hline
\hline
MgO & 1/8 & 330 & 320 \\
MgO & 1/16 & 400 & 400\\
MgO & 1/32 & 5 & 70\\
CaO & 1/16 & 72 & 83 \\
SrO & 1/8 & 82 & 66\\
SrO & 1/16 & 31 & 41\\
SrO & 1/32 & 250 & 98 \\
BaO & 1/8 &  32 & 31\\
BaO & 1/16 & 11 & 13\\
BaO & 1/32 & 280 & 230 \\
\hline
\end{tabular}
\end{center} 
\end{table}

Table \ref{tc_tab} shows how the magnetic coupling between N moments
varies both with the concentration and the different compound considered.
The data presented in the table were obtained for a calculation without considering
strong correlation effects (U= 0), but including spin-orbit effects, and
also using the Wu-Cohen GGA functional. The effect of
introducing U is to reduce the magnetic coupling, because it further localizes
the moments.
In fact, using a range of values of
U up to 8 eV, we find that, for BaO and SrO, the introduction of U as small as
1.5 eV
leads to magnetic coupling too small to determine reliably from supercell
energy differences.
The hole states no longer overlap directly, and the
materials are insulating with no means to transfer magnetic coupling
except classical dipolar coupling.
For  MgO at the higher concentrations ($x$=1/8 and $x$= 1/16), the magnetic coupling is much
higher than for the other compounds, because of the much smaller lattice parameter,
and the introduction of U reduces the size of the interaction but does not kill T$_c$.

\begin{figure}[ht]
\begin{center}  
\includegraphics[width=0.9\columnwidth,draft=false]{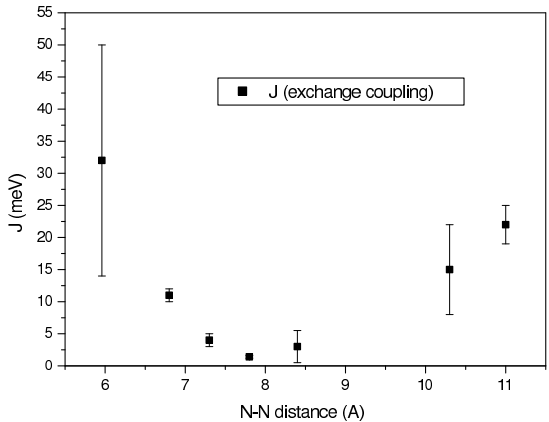}
\caption{Exchange coupling strength with respect to the N-N distance,
considering all the LSDA data from the various compounds together. ``Error'' bars
simply denote the range of values that were calculated for the same
near neighbor distance, see Table II.}\label{J_fig}
\end{center} 
\end{figure}

Putting all these LSDA data together, we can make a plot of how the magnetic interaction
strength J varies in these compounds with respect to the N-N distance, neglecting all
the differences between the various compounds we are considering.
The result can be seen in Figure \ref{J_fig}, that shows an exponential-like decay in
the strength of the magnetic coupling as the N impurity atoms are separated. There is, remarkably,
an upturn for N-N distances above 10 \AA, with values above room temperature for BaO and SrO for a
concentration around $x$ $\sim$ 3\%.
For this supercell, the magnetic impurities are located on a simple cubic sublattice at lattice constant 2a. Elfimov \textsl{et al.}\cite{elf2} have shown that the electronic coupling between N impurities in SrO is highly directional, with strong coupling along the crystalline axes (large effective p$_x$-p$_x$ coupling along the x-axis, for example).
This opens the possibility of reaching high temperature FM attainable
concentrations of N dopant in the limit U$\sim$ W.

\begin{figure}[ht]
\begin{center}
\includegraphics[width=0.95\columnwidth,draft=false]{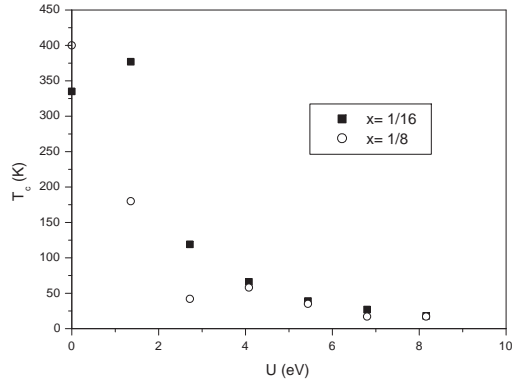}
\caption{Strength of the magnetic coupling in MgO obtained from an LSDA+U
calculation (including spin-orbit coupling) with various values of U
calculated in terms of the Curie temperature.}\label{tc_mgo_tab}
\end{center}
\end{figure}

A double-exchange-like microscopic mechanism seems to be the
best explanation of the FM coupling in these monoxides in the GGA limit, when
the Fermi level lies within the impurity band. Double
exchange is normally associated with carriers coupled to a background
of moments, with kinetic energy favored by alignment of the moments.
Electronic structure-wise, the added kinetic energy shows up as an
increased bandwidth compared to antialignment.  Indeed for the GGA
calculations and for small values of U, the system is (half) metallic
and the double exchange picture as conventionally interpreted makes
sense.  As U is increased beyond $\sim$1 eV, a gap appears, and in the 
large U regime the coupling is certain to be via direct exchange.  However, 
as shown in Fig. 3, the coupling strength (as reflected in T$_c$ in this figure) does not
appear to undergo any discontinuity at the metal-insulator transition.
The coupling (and T$_c$) does however decrease rapidly as the value
of U is increased. Having band states at the Fermi level leads also to a larger FM coupling. 
When U $>$ U$_{cr}$ = 1 eV is introduced in the calculations, an insulating state is 
obtained and magnetic exchange coupling is reduced (see Fig. \ref{tc_mgo_tab}).

In Fig. \ref{tc_mgo_tab} we illustrate more explicitly how the effect of U reduces
the Curie temperature. Strong correlation effects produce a stronger 
localization of the magnetic moments, i.e. their effects extend to fewer neighbors.
The reduced separation of overlap leads to a corresponding reduction of the 
magnetic interaction. Hence, in the
limit U $>>$ W, where this system resides for the concentrations we study, 
the magnetic coupling will not
be strong.

\section{Summary}

In this paper we have studied the formation of local moments when substitutional N impurities are 
introduced in alkaline-earth monoxides. We have calculated the coupling for a variety of 
concentrations, paying particular attention to the effect of Hubbard repulsion U (since U $>$ W 
for these systems, where W is the impurity band width). The local moments that form, 
mainly confined to the N atoms and their nearest neighbors, would lead to room temperature
ferromagnetism in the itinerant electron limit via a double-exchange-like exchange coupling
within the impurity band. 
However, in the more realistic localized electron limit U $>>$ W the coupling must be due
to direct exchange (local impurity state overlap) and becomes 
drastically reduced. Interestingly, the crossover between the double-exchange coupling and the
direct exchange as the interaction strength is increased does not lead to any discernible
anomaly in magnetic coupling strength at the metal-insulator transition.
Finally, we confirm that the impurity state is highly anisotropic, and as result magnetic
coupling is enhanced by ordering of the magnetic N impurities on a periodic (simple cubic) sublattice.

\section{Acknowledgments}

We have benefited from discussion on this topic with S. S. P. Parkin and G. A. Sawatzky.
This project was supported by DOE grant DE-FG02-04ER46111 and through interactions with
the Predictive
Capability for Strongly Correlated Systems team of the Computational
Materials Science Network.


\begin{thebibliography}{22}
\expandafter\ifx\csname natexlab\endcsname\relax\def\natexlab#1{#1}\fi
\expandafter\ifx\csname bibnamefont\endcsname\relax
  \def\bibnamefont#1{#1}\fi
\expandafter\ifx\csname bibfnamefont\endcsname\relax
  \def\bibfnamefont#1{#1}\fi
\expandafter\ifx\csname citenamefont\endcsname\relax
  \def\citenamefont#1{#1}\fi
\expandafter\ifx\csname url\endcsname\relax
  \def\url#1{\texttt{#1}}\fi
\expandafter\ifx\csname urlprefix\endcsname\relax\def\urlprefix{URL }\fi
\providecommand{\bibinfo}[2]{#2}
\providecommand{\eprint}[2][]{\url{#2}}

\bibitem[{\citenamefont{Zutic et~al.}(2004)\citenamefont{Zutic, Fabian, and
  Sarma}}]{RMP}
\bibinfo{author}{\bibfnamefont{I.}~\bibnamefont{$\check{Z}$uti\'{c}}},
  \bibinfo{author}{\bibfnamefont{J.}~\bibnamefont{Fabian}}, \bibnamefont{and}
  \bibinfo{author}{\bibfnamefont{S.~D.} \bibnamefont{Sarma}},
  \bibinfo{journal}{Rev. Mod. Phys.} \textbf{\bibinfo{volume}{76}},
  \bibinfo{pages}{323} (\bibinfo{year}{2004}).

\bibitem[{\citenamefont{Elfimov et~al.}(2002)\citenamefont{Elfimov, Yunoki, and
  Sawatzky}}]{elf1}
\bibinfo{author}{\bibfnamefont{I.~S.} \bibnamefont{Elfimov}},
  \bibinfo{author}{\bibfnamefont{S.}~\bibnamefont{Yunoki}}, \bibnamefont{and}
  \bibinfo{author}{\bibfnamefont{G.~A.} \bibnamefont{Sawatzky}},
  \bibinfo{journal}{Phys. Rev. Lett.} \textbf{\bibinfo{volume}{89}},
  \bibinfo{pages}{216403} (\bibinfo{year}{2002}).

\bibitem[{\citenamefont{Coey et~al.}(2005)\citenamefont{Coey, Venkatesan,
  Stamenov, Fitzgerald, and Dorneles}}]{vacancy}
\bibinfo{author}{\bibfnamefont{J.~M.~D.} \bibnamefont{Coey}},
  \bibinfo{author}{\bibfnamefont{M.}~\bibnamefont{Venkatesan}},
  \bibinfo{author}{\bibfnamefont{P.}~\bibnamefont{Stamenov}},
  \bibinfo{author}{\bibfnamefont{C.~B.} \bibnamefont{Fitzgerald}},
  \bibnamefont{and} \bibinfo{author}{\bibfnamefont{L.~S.}
  \bibnamefont{Dorneles}}, \bibinfo{journal}{Phys. Rev. B}
  \textbf{\bibinfo{volume}{72}}, \bibinfo{pages}{024450}
  (\bibinfo{year}{2005}).

\bibitem[{\citenamefont{Kenmochi
  et~al.}(2004{\natexlab{a}})\citenamefont{Kenmochi, Seike, Sato, Yanase, and
  Katayama-Yoshida}}]{osaka0}
\bibinfo{author}{\bibfnamefont{K.}~\bibnamefont{Kenmochi}},
  \bibinfo{author}{\bibfnamefont{M.}~\bibnamefont{Seike}},
  \bibinfo{author}{\bibfnamefont{K.}~\bibnamefont{Sato}},
  \bibinfo{author}{\bibfnamefont{A.}~\bibnamefont{Yanase}}, \bibnamefont{and}
  \bibinfo{author}{\bibfnamefont{H.}~\bibnamefont{Katayama-Yoshida}},
  \bibinfo{journal}{Jpn. J. Appl. Phys.} \textbf{\bibinfo{volume}{43}},
  \bibinfo{pages}{L934} (\bibinfo{year}{2004}{\natexlab{a}}).

\bibitem[{\citenamefont{Elfimov et~al.}(2007)\citenamefont{Elfimov, Rusydi, Hu,
  Hsich, Lin, Chen, Lian, and Sawatzky}}]{elf2}
\bibinfo{author}{\bibfnamefont{I.~S.} \bibnamefont{Elfimov}},
  \bibinfo{author}{\bibfnamefont{A.}~\bibnamefont{Rusydi}},
  \bibinfo{author}{\bibfnamefont{S.~I.}~\bibnamefont{Csiszar}},
  \bibinfo{author}{\bibfnamefont{Z.}~\bibnamefont{Hu}},
  \bibinfo{author}{\bibfnamefont{H.~H.} \bibnamefont{Hsieh}},
  \bibinfo{author}{\bibfnamefont{H.-J.} \bibnamefont{Lin}},
  \bibinfo{author}{\bibfnamefont{C.~T.} \bibnamefont{Chen}},
  \bibinfo{author}{\bibfnamefont{R.}~\bibnamefont{Liang}}, \bibnamefont{and}
  \bibinfo{author}{\bibfnamefont{G.~A.} \bibnamefont{Sawatzky}},
  \bibinfo{journal}{Phys. Rev. Lett.} \textbf{\bibinfo{volume}{98}},
  \bibinfo{pages}{137202} (\bibinfo{year}{2007}).

\bibitem[{\citenamefont{Hong et~al.}(2006)\citenamefont{Hong, Sakai, Poirot,
  and Briz\'{e}}}]{hong}
\bibinfo{author}{\bibfnamefont{N.~H.} \bibnamefont{Hong}},
  \bibinfo{author}{\bibfnamefont{J.}~\bibnamefont{Sakai}},
  \bibinfo{author}{\bibfnamefont{N.}~\bibnamefont{Poirot}}, \bibnamefont{and}
  \bibinfo{author}{\bibfnamefont{V.}~\bibnamefont{Briz\'{e}}},
  \bibinfo{journal}{Phys. Rev. B} \textbf{\bibinfo{volume}{73}},
  \bibinfo{pages}{132404} (\bibinfo{year}{2006}).

\bibitem[{\citenamefont{Kenmochi
  et~al.}(2004{\natexlab{b}})\citenamefont{Kenmochi, Seike, Sato, Yanase, and
  Katayama-Yoshida}}]{osaka}
\bibinfo{author}{\bibfnamefont{K.}~\bibnamefont{Kenmochi}},
  \bibinfo{author}{\bibfnamefont{V.~A.}~\bibnamefont{Dinh}},
  \bibinfo{author}{\bibfnamefont{K.}~\bibnamefont{Sato}},
  \bibinfo{author}{\bibfnamefont{A.}~\bibnamefont{Yanase}}, \bibnamefont{and}
  \bibinfo{author}{\bibfnamefont{H.}~\bibnamefont{Katayama-Yoshida}},
  \bibinfo{journal}{J. Phys. Soc. Jpn.} \textbf{\bibinfo{volume}{73}},
  \bibinfo{pages}{2952} (\bibinfo{year}{2004}{\natexlab{b}}).

\bibitem[{ban()}]{band_gaps}
\bibinfo{note}{A.M. Stoneham and J. Dhote, ``A compilation of crystal data for
  halides and oxides´´ (University College London, London, 2002), availabe
  online from www.cmmp.ucl.ac.uk/~ahh/research/crystal/homepa\-g\-e.\-h\-tm, and
  references therein.}

\bibitem[{\citenamefont{Lines and Bosch}(1981)}]{lines}
\bibinfo{author}{\bibfnamefont{M.~E.} \bibnamefont{Lines}} \bibnamefont{and}
  \bibinfo{author}{\bibfnamefont{M.~A.} \bibnamefont{B\"{o}sch}},
  \bibinfo{journal}{Phys. Rev. B} \textbf{\bibinfo{volume}{23}},
  \bibinfo{pages}{263} (\bibinfo{year}{1981}).

\bibitem[{\citenamefont{Labhart et~al.}(1979)\citenamefont{Labhart, Raoux,
  Kanzig, and Bosch}}]{rbo2}
\bibinfo{author}{\bibfnamefont{M.}~\bibnamefont{Labhart}},
  \bibinfo{author}{\bibfnamefont{D.}~\bibnamefont{Raoux}},
  \bibinfo{author}{\bibfnamefont{W.}~\bibnamefont{K\"{a}nzig}}, \bibnamefont{and}
  \bibinfo{author}{\bibfnamefont{M.~A.} \bibnamefont{B\"{o}sch}},
  \bibinfo{journal}{Phys. Rev. B} \textbf{\bibinfo{volume}{20}},
  \bibinfo{pages}{53} (\bibinfo{year}{1979}).

\bibitem[{\citenamefont{Winterlik et~al.}(2007)\citenamefont{Winterlik, Fecher,
  Felser, Muhle, and Jansen}}]{rb4o6}
\bibinfo{author}{\bibfnamefont{J.}~\bibnamefont{Winterlik}},
  \bibinfo{author}{\bibfnamefont{G.~H.} \bibnamefont{Fecher}},
  \bibinfo{author}{\bibfnamefont{C.}~\bibnamefont{Felser}},
  \bibinfo{author}{\bibfnamefont{C.}~\bibnamefont{Muhle}}, \bibnamefont{and}
  \bibinfo{author}{\bibfnamefont{M.}~\bibnamefont{Jansen}},
  \bibinfo{journal}{J. Am. Chem. Soc.} \textbf{\bibinfo{volume}{129}},
  \bibinfo{pages}{6990} (\bibinfo{year}{2007}).

\bibitem[{\citenamefont{Attema et~al.}(2007)\citenamefont{Attema, de~Wijs, and
  de~Groot}}]{degroot}
\bibinfo{author}{\bibfnamefont{J.~J.} \bibnamefont{Attema}},
  \bibinfo{author}{\bibfnamefont{G.~A.} \bibnamefont{de~Wijs}},
  \bibnamefont{and} \bibinfo{author}{\bibfnamefont{R.~A.}
  \bibnamefont{de~Groot}}, \bibinfo{journal}{J. Phys.: Condens. Matter}
  \textbf{\bibinfo{volume}{19}}, \bibinfo{pages}{165203}
  (\bibinfo{year}{2007}).

\bibitem[{\citenamefont{Auffermann et~al.}(2001)\citenamefont{Auffermann,
  Prots, and Kniep}}]{auffermann}
\bibinfo{author}{\bibfnamefont{G.}~\bibnamefont{Auffermann}},
  \bibinfo{author}{\bibfnamefont{Y.}~\bibnamefont{Prots}}, \bibnamefont{and}
  \bibinfo{author}{\bibfnamefont{R.}~\bibnamefont{Kniep}},
  \bibinfo{journal}{Angew. Chem. Int. Edit.} \textbf{\bibinfo{volume}{40}},
  \bibinfo{pages}{547} (\bibinfo{year}{2001}).

\bibitem[{\citenamefont{Volnianska and Bogulawski}(2008)}]{volnianska}
\bibinfo{author}{\bibfnamefont{O.}~\bibnamefont{Volnianska}} \bibnamefont{and}
  \bibinfo{author}{\bibfnamefont{P.}~\bibnamefont{Boguslawski}},
  \bibinfo{journal}{Phys. Rev. B} \textbf{\bibinfo{volume}{77}},
  \bibinfo{pages}{220403(R)} (\bibinfo{year}{2008}).

\bibitem[{\citenamefont{Pentcheva and Pickett}(2006)}]{pentcheva}
\bibinfo{author}{\bibfnamefont{R.}~\bibnamefont{Pentcheva}} \bibnamefont{and}
  \bibinfo{author}{\bibfnamefont{W.~E.} \bibnamefont{Pickett}},
  \bibinfo{journal}{Phys. Rev. B} \textbf{\bibinfo{volume}{74}},
  \bibinfo{pages}{035112} (\bibinfo{year}{2006}).

\bibitem[{\citenamefont{Brinkman et~al.}(2007)\citenamefont{Brinkman, Huijben,
  Zalk, Huijben, Zeitler, Mann, der Wiel, Rijnder, Blank, and
  Hilgenkamp}}]{magn_IF}
\bibinfo{author}{\bibfnamefont{A.}~\bibnamefont{Brinkman}},
  \bibinfo{author}{\bibfnamefont{M.}~\bibnamefont{Huijben}},
  \bibinfo{author}{\bibfnamefont{M.} \bibnamefont{Van Zalk}},
  \bibinfo{author}{\bibfnamefont{J.}~\bibnamefont{Huijben}},
  \bibinfo{author}{\bibfnamefont{U.}~\bibnamefont{Zeitler}},
  \bibinfo{author}{\bibfnamefont{J.~C.} \bibnamefont{Maan}},
  \bibinfo{author}{\bibfnamefont{W.~G.} \bibnamefont{Van der Wiel}},
  \bibinfo{author}{\bibfnamefont{G.}~\bibnamefont{Rijnders}},
  \bibinfo{author}{\bibfnamefont{D.~H.~A.} \bibnamefont{Blank}},
  \bibnamefont{and}
  \bibinfo{author}{\bibfnamefont{H.}~\bibnamefont{Hilgenkamp}},
  \bibinfo{journal}{Nat. Mater.} \textbf{\bibinfo{volume}{6}},
  \bibinfo{pages}{493} (\bibinfo{year}{2007}).

\bibitem[{\citenamefont{Hohenberg and Kohn}(1964)}]{dft}
\bibinfo{author}{\bibfnamefont{P.}~\bibnamefont{Hohenberg}} \bibnamefont{and}
  \bibinfo{author}{\bibfnamefont{W.}~\bibnamefont{Kohn}},
  \bibinfo{journal}{Phys. Rev.} \textbf{\bibinfo{volume}{136}},
  \bibinfo{pages}{B864} (\bibinfo{year}{1964}).

\bibitem[{\citenamefont{Schwarz and Blaha}(2003)}]{wien}
\bibinfo{author}{\bibfnamefont{K.}~\bibnamefont{Schwarz}} \bibnamefont{and}
  \bibinfo{author}{\bibfnamefont{P.}~\bibnamefont{Blaha}},
  \bibinfo{journal}{Comp. Mat. Sci.} \textbf{\bibinfo{volume}{28}},
  \bibinfo{pages}{259} (\bibinfo{year}{2003}).

\bibitem[{\citenamefont{Sj{\"o}stedt et~al.}(2000)\citenamefont{Sj{\"o}stedt,
  N{\"o}rdstrom, and Singh}}]{sjo}
\bibinfo{author}{\bibfnamefont{E.}~\bibnamefont{Sj{\"o}stedt}},
  \bibinfo{author}{\bibfnamefont{L.}~\bibnamefont{N{\"o}rdstrom}},
  \bibnamefont{and} \bibinfo{author}{\bibfnamefont{D.}~\bibnamefont{Singh}},
  \bibinfo{journal}{Solid State Commun.} \textbf{\bibinfo{volume}{114}},
  \bibinfo{pages}{15} (\bibinfo{year}{2000}).

\bibitem[{\citenamefont{Wu and Cohen}(2006)}]{wu_cohen}
\bibinfo{author}{\bibfnamefont{Z.}~\bibnamefont{Wu}} \bibnamefont{and}
  \bibinfo{author}{\bibfnamefont{R.~E.} \bibnamefont{Cohen}},
  \bibinfo{journal}{Phys. Rev. B} \textbf{\bibinfo{volume}{73}},
  \bibinfo{pages}{235116} (\bibinfo{year}{2006}).

\bibitem[{\citenamefont{Lichtenstein et~al.}(1995)\citenamefont{Lichtenstein,
  Anisimov, and Zaanen}}]{sic}
\bibinfo{author}{\bibfnamefont{A.~I.}~\bibnamefont{Liechtenstein}},
  \bibinfo{author}{\bibfnamefont{V.~I.}~\bibnamefont{Anisimov}}, \bibnamefont{and}
  \bibinfo{author}{\bibfnamefont{J.}~\bibnamefont{Zaanen}},
  \bibinfo{journal}{Phys.\ Rev. B} \textbf{\bibinfo{volume}{52}},
  \bibinfo{pages}{R5467} (\bibinfo{year}{1995}).

\bibitem[{\citenamefont{Ghijsen et~al.}(1988)\citenamefont{Ghijsen, Tjeng, van
  Elp, Eskes, Westerink, Sawatzky, and Czyzyk}}]{auger_Up1}
\bibinfo{author}{\bibfnamefont{J.}~\bibnamefont{Ghijsen}},
  \bibinfo{author}{\bibfnamefont{L.~H.} \bibnamefont{Tjeng}},
  \bibinfo{author}{\bibfnamefont{J.}~\bibnamefont{van Elp}},
  \bibinfo{author}{\bibfnamefont{H.}~\bibnamefont{Eskes}},
  \bibinfo{author}{\bibfnamefont{J.}~\bibnamefont{Westerink}},
  \bibinfo{author}{\bibfnamefont{G.~A.} \bibnamefont{Sawatzky}},
  \bibnamefont{and} \bibinfo{author}{\bibfnamefont{M.~T.}
  \bibnamefont{Czyzyk}}, \bibinfo{journal}{Phys. Rev. B}
  \textbf{\bibinfo{volume}{38}}, \bibinfo{pages}{11322} (\bibinfo{year}{1988}).

\end{thebibliography}

\end{document}